\documentstyle[aps,preprint]{revtex}
\begin{document}
\draft
\title{Universal Parametric Correlations of Conductance Peaks in Quantum Dots}
\author{Y. Alhassid and H. Attias}
\address{Center for Theoretical Physics, Sloane Physics Laboratory,\\
         Yale University, New Haven, CT 06520, U.S.A.}
\date {Submitted 21 February 1995}
\maketitle

\begin{abstract}
We compute the parametric correlation function of the conductance peaks in 
chaotic and weakly disordered quantum dots in the Coulomb blockade regime 
 and demonstrate its universality upon an appropriate scaling of 
the parameter. For a symmetric dot we show that this correlation function is 
affected by breaking time-reversal symmetry but is independent of  the 
 details of the channels in the external leads. We derive a new scaling  
which depends on the eigenfunctions alone and can be extracted directly 
from the conductance peak heights. Our results are in excellent 
agreement with model simulations of a disordered quantum dot.\\
\end{abstract}

\narrowtext
	
Quantum dots \cite{Ka,Mc} are semiconductor heterostructures that 
confine about 100 electrons  by an 
electrostatic potential  to isolated islands with a typical 
size of  less than a micron.   The dots can be weakly coupled via tunnel 
barriers to external leads in order to study their transport properties. 
For sufficiently low temperatures the conductance of the dot exhibits equally 
spaced peaks with increasing gate voltage, where each successive peak 
corresponds to a tunneling of a single electron into the dot. This occurs
when  the increase in the Fermi energy 
in the leads matches the energy required to charge the dot by 
one additional electron. The suppression
of tunneling between the peaks by Coulomb repulsion is known as Coulomb 
blockade. A striking feature
of these resonances is the irregular dependence of their 
amplitudes on controllable parameters, such as the shape of the dot or
 an external magnetic field. These
fluctuations have recently been accounted for by a statistical theory \cite{JSA}
based on the assumption that the Hamiltonian of the dot can be described by
random-matrix theory (RMT) due to irregularities in the confining potential
which give rise to a chaotic classical dynamics.    RMT
description is suitable also for dots with weak impurity disorder since the same
conductance distribution is obtained when the dot is modelled by a random
potential \cite{PEI}. 

Recent experiments have been probing the conductance of quantum dots 
as an external parameter  is varied. 
In this paper we study the correlation between conductance
 peak amplitudes which belong to different values of the parameter. Very 
little is known about this parametric correlator in the regime of 
isolated resonances, in contrast with the overlapping resonance regime 
characteristic of open dots where correlations versus energy and magnetic 
field  were investigated both theoretically (for a large number of channels) 
\cite{JBS,Ef} and experimentally \cite{MW}.  By casting the peak correlator in 
the framework of the Gaussian random-matrix process (GP) \cite{AA}, we 
demonstrate its universality upon an appropriate scaling of the parameter 
and  compute its universal form.  Our main results for the conductance 
correlator are the approximate expression  (\ref{9}) and its exact 
short-distance behavior (\ref{14'}).   
We  also  derive an alternative parameter 
scaling which is extracted directly from the measured conductance peaks 
 according to Eq.
 (\ref{15}). Its universal ratio to the usual level velocity
 scaling (\ref{5}) \cite{SA} is given by Eqs. (\ref{12'}) and (\ref{13}) 
for the orthogonal and unitary cases, respectively. 

We focus on the temperature regime $\Gamma\ll kT < \Delta$,
typical of experiments \cite{Ka,Mc}, where $\Delta$ is the mean resonance
 spacing.
In this regime only  one quasi-bound state  (usually the ground state) of the
 electron gas in the dot
contributes to each conductance peak \cite{MWL,Be},  whose width is
$\sim kT$. The peak amplitude that corresponds
to a level $\lambda$ is given by \cite{Be}
\begin{eqnarray}\label{1}
   G_\lambda =\frac{e^2}{h} \frac{\pi}{2 kT} \frac{\Gamma^l_\lambda
 \Gamma_\lambda^r}{\Gamma_\lambda^l + \Gamma_\lambda^r} \;,
\end{eqnarray}
where $\Gamma_\lambda^{l(r)}$  is the total decay width
from  level $\lambda$ into the left (right) lead.  
Each lead can support several open channels
so that $\Gamma_\lambda^{l(r)} = \sum_c \Gamma_{c\lambda}^{l(r)}$, where
 $\Gamma_{c\lambda}^{l(r)}$ 
is the partial decay width into channel $c$.
A suitable formalism that relates the level width to the 
dot's eigenfunction is the $R$-matrix theory \cite{LT}, according to which
$\Gamma_{c\lambda}=  \mid \gamma_{c\lambda}\mid^2$ where 
\begin{eqnarray}\label{2}
   \gamma_{c\lambda} =  \left( \hbar^2 k_c P_c / m \right)^{1/2} 
   \int  dS\Phi_c^*\Psi_\lambda  
\end{eqnarray}
is the partial width amplitude, $k_c$ is the channel momentum
 and $P_c$ is a penetration factor to tunnel  through 
the barrier. Here $\Psi_\lambda$ is the dot eigenfunction, 
$\Phi_c$ is the transverse wavefunction in the lead corresponding  to an open 
channel $c$, and the integral extends over the cross-section of the lead at 
its end attached to the dot. 

  The connection with RMT \cite{JSA} is made by assuming that the statistical
 properties of  the dot's  Hamiltonian are well described by a  
$N\times N$ random matrix $H$ taken from the appropriate Gaussian 
ensemble (GE).  The dot's eigenfunction  $\Psi_\lambda$  is related to the
$\lambda$-th eigenvector of $H$,  $\psi_{\lambda\mu}$, via the expansion
$\Psi_\lambda(\vec{r})=\sum_{\mu=1}^N \psi_{\lambda\mu}
   \rho_\mu (\vec{r})$,
where $\rho_{\mu}(\vec{r})$ are a set of solutions of Schr\"{o}dinger's
equation at energy $E_\lambda$ inside the dot.
Omitting the subscript $\lambda$, the partial width amplitude (\ref{2})
 can be written as a scalar product 
$\gamma_c=\sum_{\mu} \phi_{c\mu}^\ast\psi_\mu
   \equiv\langle\phi_c\mid\psi\rangle$, where
$\phi_{c\mu}\equiv(\hbar^2 k_c P_c /m)^{1/2}\int dS\Phi_c\rho_\mu^\ast$.
 The correlations between the $\gamma_c$ define the channel
correlation matrix $M$,  and are  given by the scalar products of the
 channel vectors
  $ M_{cc^\prime}\equiv\overline{\gamma_c^\ast\gamma_{c^\prime}}
   ={1\over N}\langle\phi_c\mid\phi_{c^\prime}\rangle$.
The channels can have any degree of  correlation among them 
and different decay widths into them,  hence $\phi_c$ are in general
 not orthogonal and  have different norms.

In order to discuss the correlation between conductances at different shapes or
external fields we use the framework of the GP,
which generalizes Dyson's GE  to describe statistical properties of
systems that depend on a parameter  \cite{AA,Wl}. 
A GP is a set of random $N \times N$ matrices $H(x)$ whose elements are
distributed at each $x$ according to the appropriate GE
with a prescribed correlation among elements at different values of $x$: 
\begin{eqnarray}\label{4}
   \overline {H_{\lambda\sigma} (x) } & = & 0 \;, \nonumber \\
   \overline{ H_{\lambda\sigma} (x) H_{\mu\nu} (x^\prime) }  & = & 
   {a^2 \over {2\beta}} f(x - x^\prime) g^{(\beta)}_{\lambda\sigma,\mu\nu} \;,
\end{eqnarray}
where 
 $g^{(\beta =1)}_{\lambda\sigma,\mu\nu} = 
  \delta_{\lambda\mu}\delta_{\sigma\nu}+
\delta_{\lambda\nu}\delta_{\sigma\mu}$  
for the  Gaussian orthogonal process (GOP)
and $g^{(\beta =2)}_{\lambda\sigma,\mu\nu} = 
     2\delta_{\lambda\nu}\delta_{\sigma\mu}$ 
for the Gaussian unitary process (GUP).
 Assuming that the leading order of $f$ is
 $f\approx 1-\kappa (x-x^\prime)^2$, we have 
shown \cite{AA} that $\kappa =\beta{{\pi^2}\over 4}{1\over N}
               \overline{(\partial E_\lambda/\partial x)^2}/\Delta^2$ 
which suggests scaling $x$ by the
RMS of the level velocity as originally discussed in \cite{SA}:
\begin{eqnarray}\label{5}
   x\rightarrow\bar{x} =
   \left[ \overline{ (\partial E_\lambda
     / \partial x} )^2  / \Delta^2\right]^{1/2} x \equiv \sqrt{D} x \;.
\end{eqnarray}
After the scaling (\ref{5}) 
$f\approx 1-\beta{{\pi^2}\over 4}(\bar{x}-\bar{x}^\prime)^2/N$
becomes independent of the non-universal quantity $\kappa$.  The parametric 
correlator for any observable, being determined by $N(1-f)$,
is then  a universal function of $\bar{x}-\bar{x}^\prime$ \cite{AA}. This
applies not only to spectral correlators \cite{SA} but also to those
involving the eigenfunctions. In particular, the conductance peak correlator
\begin{eqnarray}\label{6}
   c_G(x-x^\prime) \equiv \overline{\tilde{G}(x)\tilde{G}(x^\prime)}
   \;,\;\;\;\;\;
   \tilde{G}(x)={{G(x)-\overline{G}}\over
                 {(\overline{G^2}-\overline{G}^2)^{1/2}}}
\end{eqnarray}
becomes universal  upon the scaling (\ref{5})
for all dots characterized by the same channel correlation matrix $M$.
 
The case of a left-right symmetric dot is particularly simple since
$\Gamma^l=\Gamma^r\equiv\Gamma$ and the conductance peak correlator
 reduces to the width
correlator $\overline{\tilde{\Gamma}(x)\tilde{\Gamma}(x^\prime)}$. 
We note that the correlation matrix  $M_{cc^\prime}$  is Hermitean and positive
definite, and hence can be transformed into a diagonal form which
defines a set of  orthonormal  eigenchannel vectors $\hat{\phi}_c$.  
Expressing the
 width  in terms of  the partial widths of the normalized eigenchannels 
$\hat{\Gamma}_c=\mid\langle\hat{\phi}_c\mid\psi\rangle\mid^2$, and
using the property that the cross-channel  correlator is smaller by $1/N$
 than the autochannel correlator, we obtain in the limit of large $N$ \cite{AA1}
  \begin{eqnarray}\label{8}
   c_G(x-x^\prime)=\overline{\tilde{\hat{\Gamma}}_1(x)
\tilde{\hat{\Gamma}}_1(x^\prime)}
   \;,    
\end{eqnarray}
where $\hat{\Gamma}_1$ is the partial width of an arbitrary normalized channel.
Remarkably, the peak correlator for a symmetric dot is not only universal but
is also independent of the details of the channels in the external leads,
including their number, the rate of tunneling into each and the correlations
among them. We computed this function for both the orthogonal and the 
unitary cases from simulations of the simple GP defined by \cite{AW}
$H(x)=H_1\cos x+H_2\sin x$ where $H_{1,2}$ are independent GOE
 (GUE) matrices.
The results are presented in Fig. \ref{fig1}. We found that $c_G$ is fitted
very well
by a Lorentzian in the orthogonal case ($\beta=1$) and by a squared 
Lorentzian in the unitary case ($\beta=2$):  
\begin{eqnarray}\label{9}
   c_G(x-x^\prime)=\left[{1\over{1+(\bar{x}-\bar{x}^\prime)^2/
\alpha_\beta^2}}\right]^\beta \;,
\end{eqnarray}
where $\alpha_1=0.48\pm 0.04$  and $\alpha_2= 0.64 \pm 0.04$ 
 with a $\chi^2$ per degree of freedom of $\approx 10^{-2}$. 

We tested the GP prediction
 by studying the peak correlator in a
disordered dot modelled by a two-dimensional Anderson Hamiltonian.
We used a $27\times 27$ lattice with a cylindrical geometry and introduced a 
parametric dependence by adding a step potential of strength $x$ along the 
symmetry axis. The left and right leads are represented by arrays $A^{l(r)}$ of 
$m_1\times m_2$ lattice points with total widths given by 
$\Gamma^{l(r)}=\sum_{{\bf r}_i\in A^{l(r)}}v_i^2\mid\Psi({\bf r}_i)\mid^2$.
In Fig. \ref{fig1} we show the peak correlator for $1\times 1$ and 
$4\times 4$ leads, corresponding to leads with a 
single channel and with many correlated channels, respectively. We took $v_i=1$
but verified that different choices do not change the results, which are in
excellent agreement with the GOP prediction. In the inset we show
$\Gamma(\bar{x})/\bar{\Gamma}$ for a typical member of the GOP and for one
realization of the site energies in the Anderson model, using a single
 eigenfunction without any statistical averaging. 
As expected, this quantity exhibits irregular
oscillations with a period comparable to the decorrelation distance along 
$\bar{x}$.

In order to break  time-reversal symmetry,  we applied a constant magnetic 
field along the symmetry axis, tuned such that the cylinder encloses a flux of 
$1/4$ flux unit. The parametric dependence was
introduced either by a step potential as before or by closing the cylinder into
a torus and applying a varying magnetic field perpendicular to the constant
one.
The peak correlators in both cases are also plotted in Fig. \ref{fig1}, 
and agree very well with the GUP prediction.

Although in principle all parametric correlators become universal upon the
scaling (\ref{5}), in practice this is not always experimentally feasible.
 The energy levels probed in a quantum dot are usually
 not the excited states for a fixed number of electrons,  but the ground 
states for different  numbers of electrons.   Since the spacing between the
 conductance peaks is dominated by the charging energy 
 $e^2/C$ which is much larger than the mean spacing $\Delta$, the scaling 
factor in (\ref{5}) is difficult to measure.  It is therefore important to  
derive a scaling  that can be extracted 
directly from the conductance peak heights. For that purpose we define
$\gamma_\lambda(x) \equiv \langle\hat{\phi}\mid\psi_\lambda(x)\rangle$
and consider its derivative
\begin{eqnarray}\label{10}
   {{\partial\gamma_\lambda}\over{\partial x}}
   =\lim_{x^\prime\rightarrow x}{1\over{x^\prime-x}}
   \sum_{\mu\neq\lambda}{
   {\langle\psi_\mu\mid H(x^\prime)\mid\psi_\lambda\rangle}
   \over{E_\lambda -E_\mu} }
   \gamma_\mu \;, 
\end{eqnarray}
given by first-order perturbation
theory and where unprimed quantities are evaluated at $x$.
The RMS of $\partial\gamma_\lambda/\partial x$
is then calculated in two steps. First, we perform the average over 
$H(x^\prime)$ at fixed $H(x)$, 
  employing the conditional probability distribution \cite{AA}  for which 
$\overline{H_{\mu\lambda}(x^\prime)H_{\nu\lambda}^\ast(x^\prime)}
   =\delta_{\mu\nu}a^2(1-f^2)/\beta$.   Next we average over
$H(x)$, taking advantage of the factorization of $P\left[H(x)\right]$ into
a product of the eigenvalue and eigenfunction probability distributions. We 
obtain the following expression for the non-universal quantity $\kappa$ in the 
expansion of $f$ preceding (\ref{5}): 
\begin{eqnarray}\label{11}
   \kappa={1\over \beta {ZN}} 
          \overline{\mid{{\partial\gamma_\lambda}\over{\partial x}}\mid^2}
         /\overline{\mid\gamma_\lambda\mid^2}
\end{eqnarray}
where the constant 
$Z\equiv (a^2/\beta N) \overline{\sum_{\mu\neq\lambda}(E_\lambda-E_\mu)^{-2}}=
(4/\beta\pi^2)\int_0^\infty d\epsilon[1-Y_2(\epsilon)]/\epsilon^2$,
and $Y_2(\epsilon)$ is the two-point cluster  function.
Eq. (\ref{11}) suggests a new scaling: 

\begin{eqnarray}\label{12}
   x\rightarrow\bar{x}_r=\left( {1 \over \beta}
 \overline{\mid{{\partial\gamma_\lambda}\over{\partial x}}\mid^2}/   
  \overline{\mid\gamma_\lambda\mid^2}
     \right)^{1/2}x \equiv \sqrt{R} x \;.
\end{eqnarray}
With this scaling, $f\approx 1-(\bar{x}_r-\bar{x}_r^\prime)^2/ZN$, 
and all correlators become universal functions of $\bar{x}_r-\bar{x}_r^\prime$.
The scaling factor in (\ref{12}) can be interpreted as the RMS of 
the eigenfunction rotation rate  in analogy with the energy
level velocity in the scaling (\ref{5}).
 The quantity $Z$ above diverges for the
GOP but is finite for the GUP $ Z = 2/3$, hence the scaling (\ref{12})
 is well-defined only in the unitary case.   In this case (GUP) the ratio
 between the rotation
rate and level velocity scaling factors  is a universal constant,
obtained from comparing (\ref{11}) with the expression for $\kappa$ preceding 
(\ref{5}) 
 
\begin{eqnarray}\label{12'}
 R/D = \pi^2 /3  \;\;\;\; {\rm  (GUP)}\;.
\end{eqnarray}

The divergence of the scaling factor in (\ref{12}) for the GOP notwithstanding,
it is still possible to use this scaling if the derivative 
$\partial\gamma_\lambda/\partial x$ is replaced by  
$\Delta\gamma_\lambda/\Delta x$. This
regularization translates into a small-spacing cutoff $\delta$ in $Z$ such that
$\mid E_\lambda-E_\mu\mid/\Delta<\delta$ are excluded from the sum, resulting 
in a logarithmic divergence $Z\propto\log\delta$. For a small but finite 
$\Delta x$ we can then deduce the ratio between the two scaling factors for the 
GOP \cite{AA1}:

\begin{eqnarray}\label{13}
   {R/D}=
   -{{\pi^2}\over 6}\log(\Delta \bar{x})+Const. \;\;\;\; {\rm (GOP)}\;.
\end{eqnarray}
Relation (\ref{13}) is a universal function of $\Delta\bar{x}$,
as we demonstrate in the inset of Fig. \ref{fig1} and as is also confirmed in 
 simulations of the Anderson model.

 We now come back to the conductance peak 
correlator (\ref{8}) and extract its short-distance behavior.
We consider the quantity
$(\Delta\Gamma_\lambda)^2\equiv
 \left[\Gamma_\lambda(x^\prime)-\Gamma_\lambda (x)\right]^2$
and calculate its average in first-order perturbation theory by expanding 
$\psi_\lambda(x^\prime)$ in  $\psi_\mu(x)$.
Averaging first over $H(x^\prime)$ we obtain

\begin{eqnarray}\label{14}
   \overline{(\Delta\Gamma_\lambda)^2}
   \approx {{2a^2}\over\beta} (1-f^2) \sum_{\mu\neq\lambda}
    \overline{\left[E_\lambda -E_\mu \right]^{-2}} 
  \overline{\Gamma_\lambda \Gamma_\mu} \;, 
\end{eqnarray}
where the remaining average is over $H(x)$. Using  the GE relation
$\overline{\Gamma_\lambda \Gamma_\mu}=\left[\beta N/2(N-1)\right] 
 (\overline{\Gamma_\lambda^2}-\overline{\Gamma_\lambda}^2)$
and the scaling (\ref{12}) we find 

\begin{eqnarray}\label{14'}
c_G(x - x^\prime) \approx 1-b_\beta {R \over D} \mid \bar{x} - 
\bar{x}^\prime \mid^2 \;,
\end{eqnarray}
where $b_\beta=\beta$ and $R/D$ is given by (\ref{12'}) for the GUP and by 
(\ref{13}) for the GOP.
When compared with the leading-order behavior of the squared Lorentzian
 (\ref{9}) for the GUP case we find $\alpha_2=\sqrt{3}/\pi\approx 0.55$.
The discrepancy with the value quoted below (\ref{9})
indicates that (\ref{9}), while being a good approximation, is not exact;
indeed, higher order terms in (\ref{14}) introduce odd powers 
of $\bar{x} - \bar{x^\prime}$. 
For  the GOP, the non-analytic behavior of (\ref{14'}) at the origin indicates
that (\ref{9}) is not exact also in that case.  

 Using the perturbative expression (\ref{14}), it is possible to express
 the rotation rate scaling factor $\sqrt{R}$ directly as the RMS 
of the normalized conductance peak  velocity 
\begin{eqnarray}\label{15}
 R =  r_\beta {1 \over  \bar{G}_\lambda^2}
\overline{\left( {\partial G_\lambda \over \partial x} \right)^2} \;,
\end{eqnarray}
where $r_\beta=1/4$ and for the GOP the derivative is discrete. 
One can calculate $R$ 
from the conductance peak data according to (\ref{15})  and 
then use (\ref{12'}) or (\ref{13})  to determine $D$
and thus the scaled parameter $\bar{x}$ that leads to universal correlations. 
A semiclassical calculation of  $R$ \cite{AA1} for a ballistic electron in a 
magnetic field  ($x =B$) leads to an estimate of the correlation field
$B_c= R^{-1/2} \propto (2mE{\cal A} / \hbar^2)^{1/4} (\Phi_0 / {\cal A})$, 
where ${\cal A}$ is the area of the dot and $\Phi_0$ is the flux unit.

Finally, we computed the peak correlator for an asymmetric dot using
(\ref{1}),
this time with $\Gamma^l\neq\Gamma^r$. The left and right leads
each have their own channel correlation matrix  with
no correlation between them for a sufficiently large separation.
 Fig. \ref{fig2} displays the GP universal
predictions for $c_G$ for single-channel symmetric leads 
($\overline{\Gamma^l} = \overline{\Gamma^r}$),
and a comparison with Anderson model simulations. 
 Eq.  (\ref{9}) still provides a good fit but with $\alpha_1=0.37 \pm 0.04$ 
and $\alpha_2=0.54 \pm 0.04$.  We also find that Eq. (\ref{14'}) holds but
 with $b_1=7/4$ and $b_2=3$ \cite{AA1}.  The rotation scaling factor is
 calculated from (\ref{15}) where now $r_1=1/7$ and $r_2=5/24$ \cite{AA1}.
 For asymmetric single-channel leads 
($\overline{\Gamma^l} \neq \overline{\Gamma^r}$), $c_G$ is found to 
be intermediate between the symmetric single-channel leads correlator
and  the symmetric dot correlator.  The latter is experimentally measurable 
in an asymmetric dot whose leads are made very asymmetric, since
  the conductance peak is then approximately proportional to the dominating 
width.   For symmetric leads with many equivalent channels 
 the peak correlator also approaches the width correlator \cite{AA1}. 

In conclusion, we computed the universal conductance peak correlator
and  obtained a good
approximation (\ref{9}) of its functional form. 
We derived the rotation rate scaling  which is useful when the
 level velocity scaling factor is not measurable and  which can be directly 
calculated  from the conductance peaks data using (\ref{15}).  Since
 the submission of this Letter, our prediction (\ref{9}) has been confirmed
 experimentally for broken time-reversal symmetry \cite{Ma}.
We also remark that  the
width correlator (\ref{8},\ref{9}) is identical to the correlator of  
 wavefunction intensity at a fixed spatial point which can be measured
 in  microwave cavity experiments\cite{Sh} as a function of  shape.
This work was supported in part by the Department of Energy Grant
 DE-FG02-91ER40608. We acknowledge C.M. Marcus and A.D. Stone 
for  useful discussions.

\begin{figure}[p]

\caption{Universal form of the width correlator  (\protect \ref{8}) as a 
function of the scaled parameter  (\protect \ref{5}).
Top: GP calculations (diamonds) and their best fit to (\protect \ref{9}) 
(dashed). 
For comparison we also plot the GUP result on the left and the GOP
 result on the right  (dotted).
Insets: the universal ratio $R/D$ computed with a finite $\Delta \bar{x}$ 
using simulations of the GP (diamonds with 
statistical errors) and the Anderson model (squares and circles). The dashed
line in the left inset is the theoretical prediction (\protect\ref{13}).
Bottom: Anderson model simulations. Left: leads of 
$1\times 1$ (pluses) and $4\times 4$ (squares) points with a varying step
potential. Right: a magnetic flux of ${1 \over 4}\Phi_0$ is applied for
 $1\times 1$ leads
 with an additional varying  magnetic field (crosses) 
and $4\times 4$ leads with a varying step potential (circles).
Insets: typical width fluctuations for a single eigenfunction before the 
statistical averaging  for the GP (solid) and Anderson model (dashed).}
\label{fig1}

\vspace{5 mm}

\caption{Universal form of the peak correlator in asymmetric dots  with
 symmetric single-channel leads for  both orthogonal and unitary symmetries.
 The GP predictions (dashed)  are compared with  Anderson
model simulations  (pluses and crosses as in Fig. 1).
Shown  by dotted lines is the width correlator which also describes the
 conductance correlator for highly asymmetric leads.}
\label{fig2}
\end{figure}


\begin{references}
\bibitem{Ka} M.A. Kastner, Rev. Mod. Phys. {\bf 64}, 849 (1992).
\bibitem{Mc} P.L. McEuen {\it et al.} , Phys. Rev. Lett. {\bf 66}, 1929 (1991).
\bibitem{JSA} R.A. Jalabert, A.D. Stone, and Y. Alhassid, Phys. Rev. Lett. 
   {\bf 68}, 3468 (1992).
\bibitem{PEI} V.N. Prigodin, K.B. Efetov and S. Iida, Phys. Rev. Lett. {\bf 71},
 1230 (1993); E.R. Mucciolo, V.N. Prigodin, and B.L. Altshuler, Phys. Rev. B
 {\bf 51}, 1714(1995).
\bibitem{JBS} R.A. Jalabert, H.U. Baranger and A.D. Stone, Phys. Rev.
Lett. {\bf 65},  2442 (1990);   H.U. Baranger, R.A. Jalabert and A.D. Stone,
Chaos {\bf 3}, 665 (1993).
\bibitem{Ef} K.B. Efetov, Phys. Rev. Lett. {\bf 74}, 2299 (1995).
\bibitem{MW} C.M. Marcus, R.M. Westervelt, P.F. Hopkins and
 A.C. Gossard,  Phys. Rev B {\bf 48}, 2460 (1993); 
 I.H. Chan, R.M. Clarke, C.M. Marcus, K. Campman and A.C. Gossard, 
Phys. Rev. Lett. {\bf 74}, 3876 (1995).
\bibitem{AA} Y. Alhassid and  H. Attias,  Phys. Rev. Lett.  {\bf 74},
 4635 (1995).
\bibitem{SA} B.D. Simons and B.L. Altshuler, Phys. Rev. Lett. {\bf 70}, 4063 
   (1993); Phys. Rev. B {\bf 48}, 5422 (1993).
\bibitem{MWL} Y. Meir, N. Wingreen and P.A. Lee, Phys. Rev. Lett.
 {\bf 66}, 3048 (1991).
\bibitem{Be}  C.W.J. Beenakker, Phys. Rev. B {\bf 44}, 1646 (1991).
\bibitem{LT} A.M. Lane and R.G. Thomas, Rev. Mod. Phys. {\bf 30}, 257 (1958).
\bibitem{Wl} M. Wilkinson, J. Phys. A {\bf 22}, 2795 (1989).
\bibitem{AA1} Y. Alhassid and H. Attias, to be published.
\bibitem{AW} E.J. Austin and M.Wilkinson, Nonlinearity {\bf 5}, 1137 (1992).
\bibitem{Ma} J.A. Folk, S.R. Patel, S.F. Godijn, A.G. Huibers,
S.M. Cronenwett, C.M. Marcus, K. Campman and A.C. Gossard,
preprint  (September 1995).
\bibitem{Sh}  S. Sridhar, D. Hogenboom and B. Willemsen, J. Stat. Phys. 
{\bf 68}, 239 (1992).
\end{references}
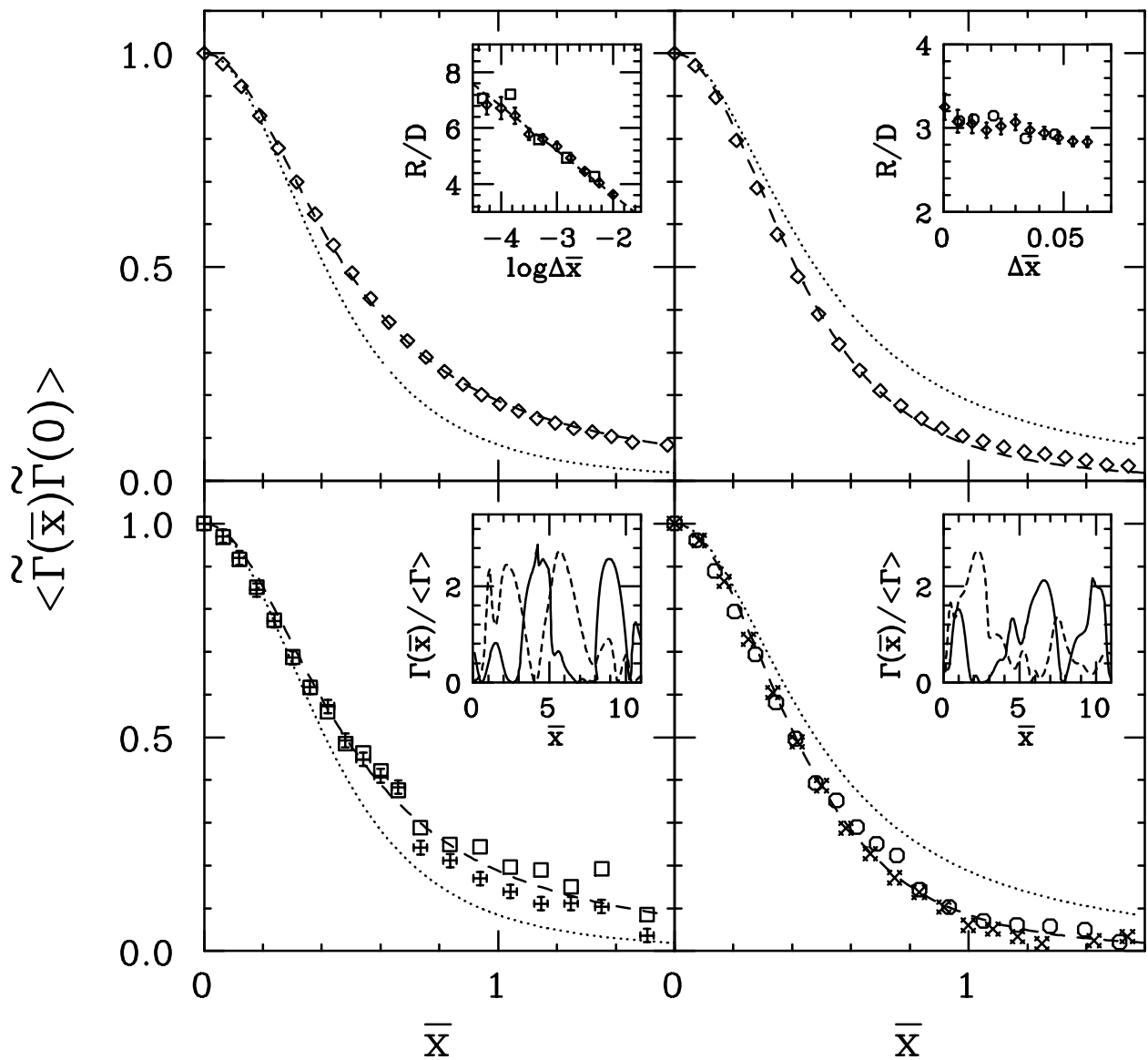
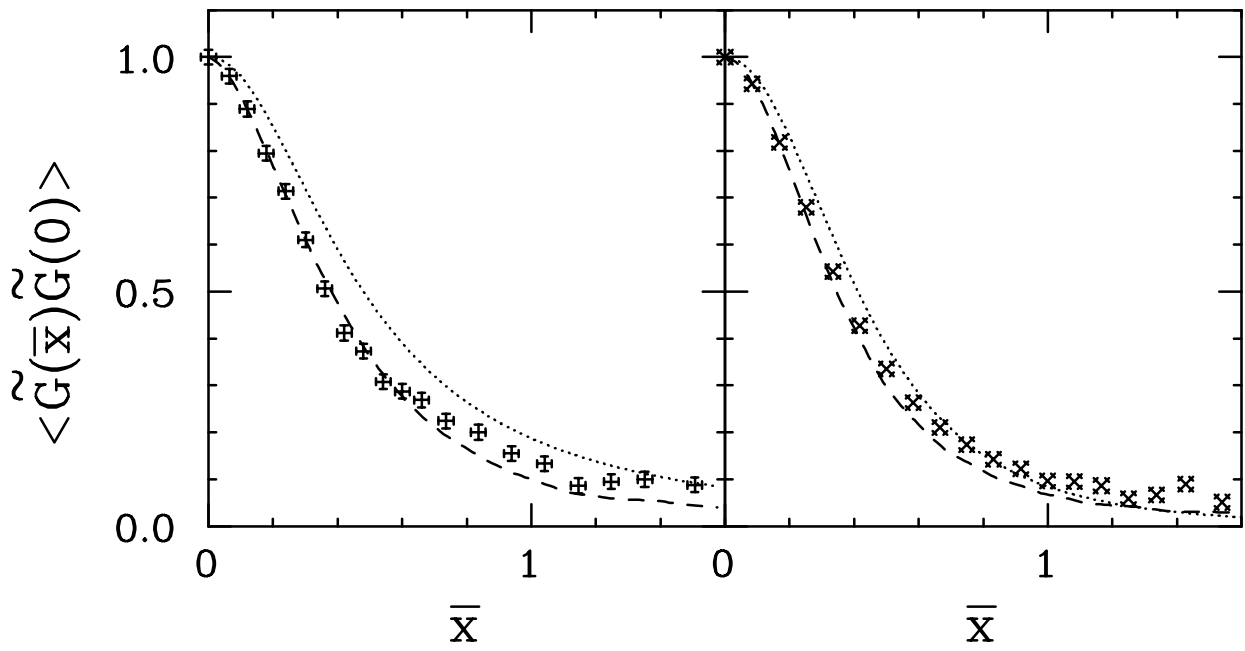
\end{document}